\documentclass{emulateapj}   

\usepackage{graphicx,amsmath}

\slugcomment{Submitted to ApJL}

\shorttitle{GSMT Lyot Coronagraphy}
\shortauthors{Sivaramakrishnan and Yaitskova}

\eqsecnum
\tightenlines

 \received{25 November 2004}

\newcommand \cf{cf.}
\newcommand \eg {{\it e.g., }}

\newcommand \ie{{\it i.e.,}}

\newcommand \shah 	       {\rm III}			
\newcommand \sinc 	{\,\,  {\rm sinc}}		    
\newcommand \eq 	{\,=\,}						
\newcommand \psfc{${\rm PSF}_c$}

\shortauthors{Sivaramakrishnan and Yaitskova }
\shorttitle{Coronagraphy on GSMTs}

\begin{document}

\title { Lyot Coronagraphy on \\ Giant Segmented-Mirror Telescopes}

\author{Anand Sivaramakrishnan} 
        \affil{Space Telescope Science Institute\\
               3700 San Martin Drive, Baltimore, MD 21218, U.S.A.}
		\and

\author{Natalia Yaitskova} 
	\affil{European Southern Observatory\\
               Karl Schwarzschildstr. 2, D-85748 Garching bei Muenchen, Germany}
                \email{anand@stsci.edu, nyaitsko@eso.org}

\begin{abstract}
We present a study of Lyot style 
(\ie\ classical, band-limited, and Gaussian occulter) coronagraphy
on extremely large, highly-segmented telescopes.
We show that while increased telescope diameter is always
an advantage for high dynamic range science (assuming wavefront
errors have been corrected sufficiently well), segmentation
itself sets a limit on the performance of Lyot coronagraphs.
Diffraction from inter-segment gaps sets a floor to the achievable
extinction of on-axis starlight with Lyot coronagraphy.
We derive an analytical expression for the manner in which
coronagraphic suppression of an on-axis source decreases
with increasing gap size when the segments are placed in a spatially periodic
array over the telescope aperture, regardless of the details of the
arrangement.  A simple Lyot stop masking out pupil edges produces good extinction of
the central peak in the point-spread function (PSF), but leaves satellite
images caused by inter-segment gaps essentially unaffected.  Masking out the bright 
segment gaps in the Lyot plane with a reticulated mask reduces the satellite images'
intensity to a contrast of $ 5 \times 10^{-9}$ on a 30\ m telescope with 10\ mm gaps,
at the expense of an increase in the brightness of the central peak.
The morphology of interesting targets will dictate which Lyot stop geometry 
is preferable: the reticulated Lyot stop produces a
conveniently uni-modal PSF, whereas a simple Lyot stop produces
an extended array of satellite spots.  
A cryogenic reticulate Lyot stop
will also benefit both direct and coronagraphic mid-IR imaging.
\end{abstract}
	    
\section{Introduction}\label{introduction}

Current conventional wisdom deems it to be easier to construct segmented
30--100~m diameter ground-based, orbiting, or lunar telescopes than monolithic ones.
We inspect the consequences of the segmented design choice for diffraction-limited
high dynamic range Lyot style coronagraphy, which is one of the techniques
used for faint companion searches and extrasolar planetary and disk science
\citep{Lyot39, Golimowski95, Nakajima95, Sivaramakrishnan01, Aime03bb, Aime04bb, Oppenheimer03RST}.
Just as high dynamic range coronagraphy is very sensitive to the amount
of residual aberration in a system \citep{Angel94}, we show that
Lyot coronagraphy is similarly sensitive to the amount of non-reflective area
in a highly segmented telescope. 
Other forms of high dynamic range imaging (such as phase mask coronagraphy,
interfero-coronagraphy, or apodized and shaped pupils) may be less
affected by mirror segmentation, but the simplicity and robustness of
Lyot coronagraphy
makes this topic relevant to the planning of astronomical instrumentation
on future telescopes.
Giant segmented-mirror telescope (GSMT) design must take the interaction
between aperture geometry and coronagraphic performance into account
if planetary companion and other high dynamic range
science is a goal for the telescope.

We describe the point-spread function (PSF) of a Lyot coronagraph on a
perfect segmented mirror telescope using the techniques of Fourier optics
in the highly segmented regime.
There are two scales of relevance in this problem.
A Lyot coronagraph with an occulting spot $s$ resolution
elements in diameter has a natural scale length  $D/s$ in its
Lyot pupil plane (as we explain in section \ref{corsummary}),
where $D$ is the aperture diameter
(see \eg\ \citet{Sivaramakrishnan01} and references therein 
for an introductory exposition of Lyot coronagraphy).
When the segment spacing $b$ is considerably smaller than this scale,
it is easy to calculate the on-axis coronagraphic PSF (\psfc).
When the inequality  $b \ll D/s$ holds we describe the telescope
as being `highly segmented', from a coronagraphic perspective.
\citet{Makidon00, Sivaramakrishnan01} demonstrated that high dynamic range,
stellar Lyot coronagraphy on unapodized apertures is only interesting when
the spot diameter exceeds about 4 resolution elements.
A $0\arcsec.034$ occulting spot on the 36-segment Keck
aperture ($s=4$ in $H$-band) does not constitute a highly
segmented telescope from the coronagraphic point of view ---
Lyot pupil diffraction is on a scale of $D/4$, and segments are $D/7$ across.
There is no clear separation of these two scales.
\citep{Sivaramakrishnan04xaopi} treat the Keck case.

Multiple-occulting spot coronagraphs that occult all the satellite PSFs in
the first image plane have been studied by 
\citet{ASL01} in the one-dimensional, sparse aperture array case,
with a dispersed image plane to address chromaticity problems.
This approach looks unpromising for almost-filled GSMT apertures 
with anything but narrow bandpass filters.

In this paper we obtain approximate expressions for on-axis 
\psfc 's using analytical
methods that are valid for Lyot style coronagraphs on GSMTs.  
This enables us to assess the dynamic range accessible to simple
coronagraphs on GSMTs, and engenders a sound physical understanding
of Lyot coronagraphy on GSMTs being planned today
(\eg\ \citet{Nelson00, Dierickx00}).
We do not treat phase mask or apodized pupil Lyot coronagraphy here.

\section{ The Aperture Illumination Function} \label{segail}

In this introductory paper we describe the aperture of a segmented
mirror telescope in a somewhat more general way than is typically
done for particular telescope design studies such as
\citet{Chanan98, Yaitskova03}.   Using a
more general formalism enables us to identify and quantify 
limitations of Lyot coronagraphs on highly segmented
apertures with arbitrary but periodic segment geometry analytically.

The unsegmented full aperture of the telescope is described by a 
function $A(x)$, which, for unapodized apertures, is unity over
the aperture and zero elsewhere.  Here $x=(u,v)$ is the location in
the aperture, in units of the wavelength of the light.
We develop the expression for the monochromatic PSF
with perfect optics first, and then calculate \psfc
(we assume that the scalar wave approximation and Fourier optics provide
an adequate description of PSF formation).

Segment gaps are described by some lattice-like
function $L(x)$, which takes the value of
unity where gaps exist, and zero over
the segments themselves.
The function $L(x)$ possesses whatever two-dimensional
crystallographic symmetry suits the problem.  
We illustrate our arguments with the simple case of
a circular aperture with square segments, although our 
results hold for the general case of arbitrary spatially-periodic
segmentation on obscured apertures.

The aperture illumination function --- the complex amplitude in the aperture
in response to a unit (field rather than power) strength incoming wave ---
for the segmented aperture, can be written as
	\begin{equation} \label{Aseg}
	A_s(x) \eq A(x)~(1 -  L(x)).
\end{equation}
In the particular case of tetrad symmetry describing
square segments,
	\begin{equation} \label{Lsquare}
	L(x) = L_1(u; b, e) + L_1(v; b, e) - L_1(u; b, e) ~ L_1(v; b, e)
\end{equation}
(where
	\begin{equation} \label{L1}
	L_1(u; b, e) \equiv \frac{1}{b}  \shah(u/b) * \Pi(u/e),
\end{equation}
and $*$ denotes convolution).
We make use of Fourier analytic techniques and results
which can be found in \eg\ \citet{Bracewell}.
Here $b$ is the segment  spatial periodicity and $e$ the 
inter-segment gap width (all pupil plane dimensions
are stated in units of the wavelength of the monochromatic
light being considered).   
The one-dimensional shah or comb function $\shah$ is defined by 
	\begin{equation} \label{shah}
	\shah(u) \equiv \sum_{n = -\infty}^{\infty} \delta(u - n),
\end{equation}
where $\delta(u)$ is the Dirac delta function.
Normalization by $1 / b$ is required to ensure 
the correct `impulse strength'
in each delta function in equation (\ref{L1}).
The top-hat function $\Pi$ is given by
    \begin{equation} \label{tophatdef}
	\begin{array}{ll} \Pi(x) \eq 1 & {\rm for\ } |x| < 1/2, \\
	                  \Pi(x) \eq 0   & {\rm elsewhere}.
	\end{array}
    \end{equation}

A scientifically relevant example
of a highly segmented aperture coronagraph might be a
30~m telescope with 1~m segment sizes and 5--10~mm gaps, 
with a hard-edged or apodized occulting spot size of
$6 \lambda/D$ at the $H$-band central wavelength ($1.63 \micron$).

\section{ A Summary of Lyot Coronagraphy } \label{corsummary}

In order to understand coronagraphy it is essential to
have a thorough understanding of the
field strength (\ie\ the wave's complex amplitude)
in the Lyot pupil plane. \citet{Sivaramakrishnan01}
discuss the fundamentals of Lyot coronagraphy in detail:
here we summarize a few key points which we use in our analysis
of segmented mirror coronagraphy.

Given an aperture illumination 
function $A$ (no matter whether it is segmented, filled, obstructed,
or apodized), the `amplitude spread function' or ASF (the field strength
in the first image plane) is given by the Fourier transform of $A$,
$a(\xi, \eta)$.  As a rule we switch the case of a function to denote
its Fourier transform: $A$ and $a$ are Fourier transform pair.
In a Lyot coronagraph this field is then modified by passage through
an occulting mask with a transmission factor $m(\xi, \eta)$ in the image
plane.  The field strength (\cf\  the intensity) is $ma$ after the mask.
It is convenient to write the transmission function $m$ as $1 - w$:
typically $w(0,0) = 1$, \ie\ the mask is opaque at the center.
When the entrance pupil of the coronagraph is re-imaged
after the bright, on-axis stellar image has been occulted by the mask,
the pupil field in this Lyot pupil plane is the Fourier transform of
$a(1-w)$, or $A - A*W$, where W is the Fourier transform
of the `mask shape function' $w$.  

As a concrete example, let us consider a hard-edged mask $6\lambda/D$
radians in diameter, which corresponds to a $w$ which is unity within a 
disk 6 resolution elements across. $W$, whose domain is the Lyot
pupil plane, is the well-known Airy pattern with a scale size of $D/6$.
Since $W$ has unit area (because $w(0,0) = 1$),
in the interior of the Lyot pupil the field strength is approximately zero.
Various forms of `band-limited' coronagraphs \citep{Kuchner02}, which have
mask shape functions which are band-limited in the Fourier sense
(for example,
\\  $w = \sinc^2(\xi D/6) ~ \sinc^2(\eta D/6)$)
produce Lyot pupil interior fields which are identically zero
on monolithic, filled apertures.

One measure of Lyot coronagraph mask width is the
equivalent width of the mask shape function
(\eg\ \citet{Bracewell}).  This is a measure
of the occulting power of the mask given a uniform intensity in the image plane.
The wider the mask width in image space, the narrower the edge effects are
in the Lyot plane, and the larger the zero or low field strength area
in the Lyot plane interior is
(see \citet{Lloyd05} for further details, and a comparison
of hard-edged and graded image plane mask coronagraphs).

The distance $D/s$ is the most important length scale
in the pupil plane of a Lyot coronagraph.

\section{ The Lyot Pupil Plane Field } \label{lyotpupilfield}

We can now estimate how much energy segment gaps diffract into the pupil.
In a Lyot coronagraph on a segmented telescope,
the field in the Lyot plane is given by
						\begin{equation} \label{lyotfield}
          E_{Lyot} \eq A_s(x,y) - A_s(x,y)~*~W(x,y)
     					\end{equation}
where $A_s$ is the aperture illumination function from equation (\ref{Aseg}),
and $W$ is the Fourier transform of the mask shape function $w$ which describes
the occulting mask's departure from complete transparency (see Figure 1, left panel).
Equation \ref{lyotfield}  expands to
						\begin{equation} \label{lyotfieldexpand}
          E_{Lyot} \eq A - A L - W*A + W*(A L).
     					\end{equation}
In the interior of the Lyot plane the quantity $A - W*A$ 
(which is the field strength for the coronagraphic ASF
of a monolithic mirror telescope with the same overall aperture
geometry as our segmented telescope, but without segment gaps)
is negligibly small or zero, depending on the coronagraph design. 
Dropping this term in the interior of the Lyot pupil, we obtain
						\begin{equation} \label{lyotfieldinterior}
          E_{Lyot\ Interior} \eq - A L + W*(A L).
     					\end{equation}

The equivalent width of the function $W$ is $D/s$. 
For $D  \sim$  30\ m, and a typical segment size $b \sim$ 1\ m,
the condition $ D/s \gg b$ holds  (in \citet{YS05} we show that 
a less restrictive condition,  $D/s  >\  \sim 3b$,
is in fact good enough for the following calculation to be valid).
In the highly-segmented telescope case this inequality holds, and the cited work 
demonstrates that the convolution $W*(A L)$ does not depend on the segmentation
geometry, but only on the ratio of the gap width to the segmentation
periodicity, $g = e/b$:

						\begin{equation} \label{lyotfieldlevel}
          W*(A L) \eq 1-(1-g)^2 \simeq 2g.
     					\end{equation}
An unapodized, unsegmented, undersized Lyot stop with an aperture function
$A'(x)$ is used to remove the bright annulus of light on the borders of the
Lyot pupil (Figure 1, middle panel).
We note that $A'A = A'$,  \ie\ the Lyot pupil on this monolithic
mirror telescope is contained within the aperture itself.

We also introduce a reticulated mask that blocks out the bright 
segment gaps in the Lyot pupil (Figure 1, right panel).
This corresponds to a mask function $1 - L$.  Our Lyot stop 
is therefore written as  $A'(1 - L)$.
The field immediately after our Lyot stop then becomes
						\begin{equation} \label{lyotfieldstopped}
          E_{Lyot\ Stop} \eq A'(1 - L)(W*(A L)) \eq 2g A'(1 - L).
     					\end{equation}
Comparing this equation with the expression in equation (\ref{Aseg}) for the
segmented aperture illumination itself, we notice that 
$E_{Lyot\ Stop}$ coincides with the complex amplitude for the segmented
telescope with the field strength reduced by factor $2g$ (in the interior
of the Lyot stop).
If the initial illumination of the aperture is uniform, and there are no
residual phase errors (\ie\ the segments are co-phased, and
wavefront correction by adaptive optics is perfect), we can apply
the techniques of \citet{Yaitskova03} to describe \psfc.

\section{ The Coronagraphic PSF} \label{segcorPSF}

The Lyot field described by equation (\ref{lyotfieldstopped})
produces a segmented telescope coronagraphic ASF
						\begin{equation} \label{segcorpsf}
          a_{sc} \eq a'~*~(^2\delta - l)~*~(w~(a~*~l))
     					\end{equation}
($a'$ being the ASF corresponding to the unsegmented Lyot stop $A'$,
$^2\delta$ the two-dimensional Dirac delta function, and
$l$ the Fourier transform of $L$).

The on-axis coronagraphic image, ${\rm PSF}_c$
(Figure~2, right), which results
from using a reticulated Lyot stop that masks out the 
segment gaps, is reduced by factor $4g^2$ from the PSF of
the segmented telescope (Figure~2, left) on which the Lyot coronagraph is applied.
The PSF of the segmented telescope with gaps (Figure~2, left) contains
higher-order `satellite' diffraction peaks (in addition to the central dominant peak)
located on a grid with an angular periodicity that is inversely
proportional to the segment center-to-center separation
(\ie\ the segment spatial periodicity).
In ${\rm PSF}_c$ there is also a broadening of the central and higher-order 
diffraction peaks due to the undersized Lyot stop diameter relative to the
original entrance aperture diameter.
The intensity of the higher order \psfc
diffraction peaks with the reticulated Lyot stop is
down a factor of $4g^2$ from those same satellite peaks 
without the reticulated Lyot stop.
The intensity of the central peak of the PSF before the coronagraph
(the Strehl ratio, if a monolithic aperture is the reference unaberrated PSF)
is $(1-g)^4 \simeq 1-4g$.
Hence the intensity of the central peak after the coronagraph is 
$(1-4g)4g^2 \simeq 4g^2$ relative to the monolithic aperture's PSF peak.
For a 10\ mm inter-segment gap, the value of this expression is
$4*(0.01/1.5)^2 = 2 \times 10^{-4}$.
The intensity
of the brightest peaks for a hexagonally segmented aperture
is $0.68g^2$ relative to the peak of the unsegmented aperture's PSF.

On a coronagraph with a matched reticulated Lyot stop, this ratio is
$0.68g^2*4g^2 = 2.7g^4$. 
For the same example as before (a 30\ m telescope with $1.5\ {\rm m}$
segments and a 10\ mm inter-segment gap),  $2.7g^4 = 5 \times 10^{-9}$.
Analogous estimates can be made for the diffraction peaks of any order.
The intensity around the peaks can also be obtained as the intensity of the
PSF for the segmented telescope, but again,  reduced by the factor $4g^2$.

The coronagraphic PSF of a segmented telescope is $a_{sc}~a^*_{sc}$.
A detailed analysis of the contrast and morphology of coronagraphic PSFs
on telescopes with hexagonal segmentation will be presented in \citet{YS05}.

As Figure \ref{images} shows, the coronagraphic PSF with the reticulated Lyot stop
has a central core and faint satellite spots.  This PSF is suitable for 
studying extended objects such as protoplanetary disks.  In contrast, using a simple
Lyot stop creates better suppression of the core of the PSF, but transmits the 
satellite spots caused by primary mirror segmentation.

\acknowledgements
We acknowledge helpful discussions with R. Soummer.
AS acknowledges support from 
the National Science Foundation
Science and Technology Center for Adaptive Optics, managed by the
University of California at Santa Cruz under cooperative
agreement No.  AST-9876783.
NY performed this work under the support of the European \
Extremely Large Telescope Design Study (Framework 6).

\bibliographystyle{apj}
\bibliography{ms}

\begin{figure}
	\epsscale{0.80}
	\plotone{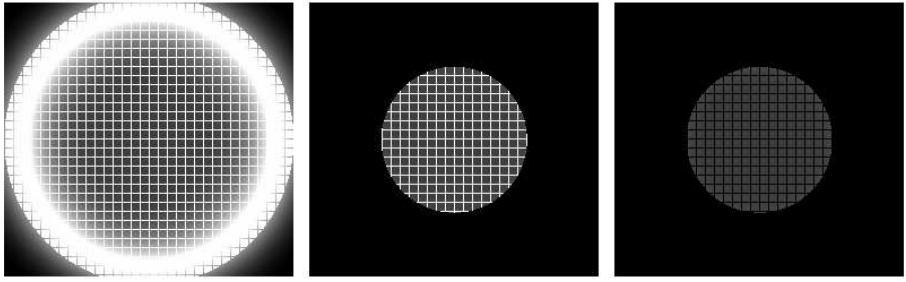}
	\caption{
		Lyot pupil plane intensities in a band-limited Lyot coronagraph (BLC)
		on a segmented aperture telescope, shown on the same linear stretch.
		Left: the BLC Lyot pupil on a square-segment aperture.
		There are 32 segments across the 512-pixel pupil diameter.
		Segment gaps are two pixels wide.
		The occulting mask transmission profile is 
		$1 - {\rm jinc}^2 kr$, where the first zero of the ${\rm jinc}$
		occurs at $16\lambda/D$.  Without segment gaps the inner circle,
		with a diameter D/2, would be completely dark.
		Middle: a simple circular Lyot stop removes diffracted light around the edge
		of the pupil, leaving bright segment gaps and a low plateau
		of illumination over clear aperture.  Right: a Lyot stop with the 
		bright segment gaps blocked by a reticulated stop in the interior.
		A cryogenic reticulated Lyot stop
		will benefit both direct and coronagraphic mid-IR imaging.
	}
	\label{pupils}
\end{figure}

     \begin{figure}
         \epsscale{0.80}
         \plotone{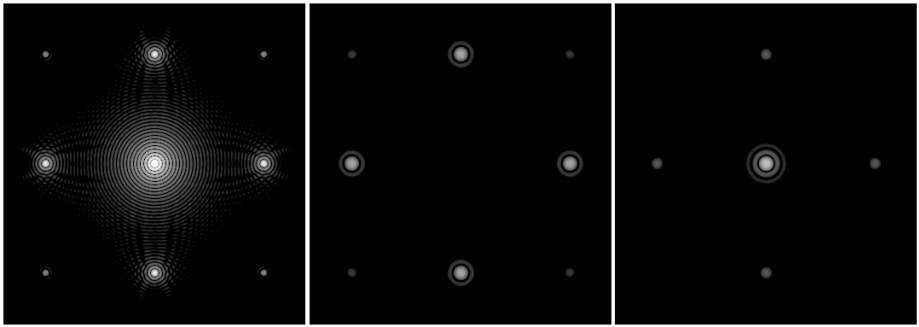}
         \caption{
		Direct and coronagraphic images 
		(shown on the same logarithmic stretch)
		from the coronagraph whose Lyot pupil is shown in Figure \ref{pupils}.
		Left: direct image showing a bright core and a grid of satellite images
		typical of segmented apertures.  The satellite images along the principal 
		symmetry axes are brighter than the other satellite images, by a factor 
		of the square of the ratio of gap width to the segment periodicity.
		Middle: coronagraphic image with the simple Lyot stop shown in the
		middle panel of Figure \ref{pupils}.  The central peak has been suppressed
		almost completely by the perfect coronagraph.  
		Right: Masking out the bright inter-segment gaps in the Lyot pupil 
		(right panel, Figure \ref{pupils})
		produces significant reduction in the satellite images' brightness,
		but creates a bright central image.  Such a unimodal on-axis PSF is suited to 
		disk studies around AO targets.
        }
\label{images}
\end{figure}

\end{document}